# Mechanism and Prevention of Hyperbaric Oxygen Convulsions

Ondrej Groborz,[a,b,*] Ludek Sefc,[c] and Petr Marsalek[d,e,*]

[a]Institute of Organic Chemistry and Biochemistry, Czech Academy of Sciences, Prague, Czech Republic
[b]Institute of Biophysics and Informatics, First Faculty of Medicine, Charles University, Prague, Czech Republic
[c]Center for Advanced Preclinical Imaging (CAPI), First Faculty of Medicine, Charles University, Prague, Czech Republic
[d]Faculty of Biomedical Engineering, Czech Technical University in Prague, Czech Republic
[e]Institute of Pathological Physiology, First Faculty of Medicine, Charles University, Prague, Czech Republic

*Corresponding authors: Ondrej Groborz, Petr Marsalek, e-mails [Ondrej.Groborz/ Petr.Marsalek]@lf1.cuni.cz

## Abstract

Hyperbaric oxygen therapy (HBOT) proves vital in saving lives by elevating the partial pressure of oxygen ($pO_2$). However, HBOT may also have toxic effects, including lung and retinal damage (peripheral HBOT toxicity), muscle spasms and violent myoclonic convulsions (CNS HBOT toxicity), which may even lead to death if left untreated. Despite the severity of the toxic effects of HBOT, their mechanism is only poorly understood to date. This lack of understanding the underlying mechanism hinders the development of new, effective therapies and preventive strategies to supress HBOT toxicity. Herein, we provide evidence that (1) increased $pO_2$ increases the content of reactive oxygen species (ROS) in tissues, which causes peripheral HBOT toxicity and contributes to CNS toxicity by irreversibly altering cell receptors. Moreover, (2) increased ROS concentration in brain lowers activity of glutamic decarboxylase (GD), which lowers concentrations of inhibitory neurotransmitter γ-aminobutyric acid (GABA), thereby contributing to the onset of HBOT-derived convulsions. At last, we provide long overlooked evidence that (3) elevated ambient pressure directly inhibits $GABA_A$ and glycine receptors, thereby leading to the rapid onset of HBOT-derived convulsions. We show that only a combination of these three mechanisms (1 + 2 + 3) are needed to explain most phenomena seen in HBOT toxicity (especially in CNS toxicity). Based on these proposed intertwined mechanisms, we propose administering antioxidants (lowering ROS concentrations), pyridoxine (restoring GD activity), and low doses of sedatives/ anaesthetics (reversing inhibitory effects of pressure on $GABA_A$ and glycine receptors) before routine hyperbaric oxygen therapies and deep-sea diving to prevent the HBOT toxicity.

## Introduction

Oxygen therapy (OT) encompasses a set of medical procedures for increasing the partial pressure of oxygen ($pO_2$) in inhaled gas, thereby increasing the supply of life-sustaining oxygen to the body. OT is indicated for numerous conditions, *e.g.*, lung disease and alveolar hypoventilation, poor tissue perfusion, carbon monoxide poisoning, gas gangrene, decompression illness, and arterial gas embolism, among others. [1,2] When a high oxygen supply is required, oxygen can be delivered in a hyperbaric chamber (hyperbaric oxygen therapy, HBOT), which further raises $pO_2$, facilitating oxygen uptake by the body. These forms of OT have become part of clinical routine and have helped save countless lives. [1,2]

HBOT is usually well tolerated by patients; the most common side effects include relatively rare traumas in the ear and sinuses caused by pressure changes. [3,4] However, HBOT may also trigger acute oxygen toxicity, with major undesirable effects.[1,3,5] Prolonged exposure to HBOT may increase the concentrations of reactive oxygen species (ROS) in tissues, [6] damaging the lungs and retinas, among other organs (Lorrain Smith effect, peripheral toxicity). [1,4,7–9] This peripheral toxicity (also called chronic toxicity[1]) of HBOT depends on the time of exposure, [1] $pO_2$ (onsets usually at 50 kPa[1]), and total gas pressure [8]. Moreover, HBOT can affect the central nervous system (CNS) and cause confusion, headache, nausea, tremors, loss of consciousness, muscular twitching, quickly progressing to painful muscle spasms and life-threatening violent tonic-clonic (*grand mal*) seizures (Paul Bert effect, CNS toxicity).[1,5,7,10]

These hyperbaric convulsions occur stochastically (**Figure 1**),[5] usually without prodromal symptoms, and thus cannot be safely predicted in advance. [1,4,11] Furthermore, susceptibility towards hyperbaric convulsions increases with oxygen consumption (such as physical activity,[5] high or low ambient temperature, [5] increased levels of cortisol, [1] thyroid hormones, [1] and epinephrine, [1] among others). Furthermore, susceptibility towards hyperbaric convulsions is high in CO poisoning patients, [12,13] hypercapnia and acidosis,[1] and other external and internal factors.[1][a] As a result, susceptibility to hyperbaric convulsions greatly varies over time (**Figure 1A**), even when external conditions are seemingly identical.[5] Nevertheless, during HBOT, increasing the total pressure and oxygen concentration increases the likelihood of hyperbaric convulsions and accelerates their onset (**Figure 1B**).[b],[5,12,14] Under moderate pressure (240 to 300 kPa), some HBOT centres report 0.008 to 0.7% incidence of hyperbaric convulsions (but 0.3 to 4.7% in CO poisoning patients[12]).[13] Yet, even under low pressure (below 210 kPa), approximately 0.01 to 0.06% patients reportedly develop hyperbaric convulsions during standard HBOT. [13,17]

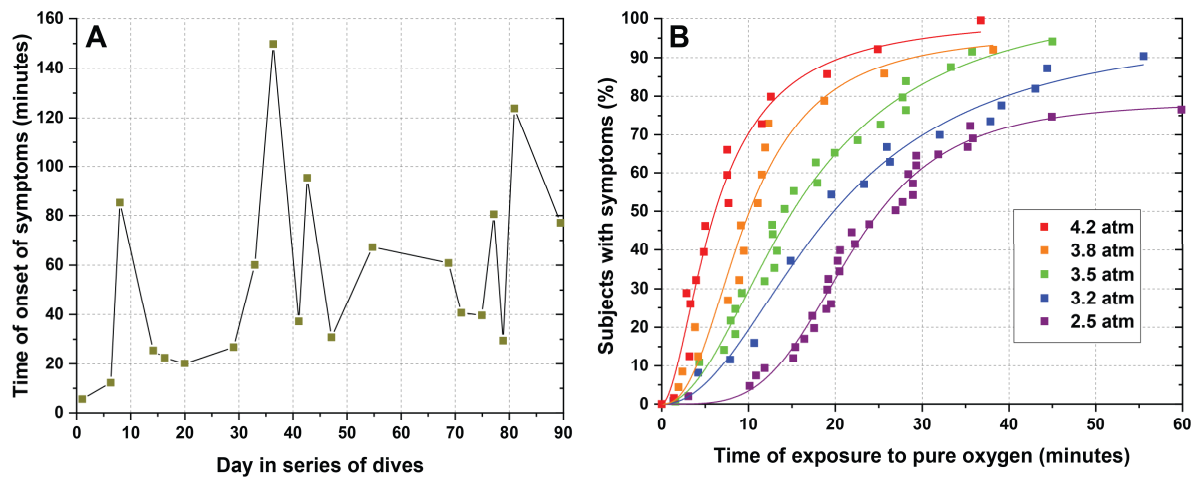

**Figure 1**. (A) Time of exposure of a single diver to high-pressure oxygen (3.2 atm.) until symptoms of hyperbaric toxicity occurred; the diver experienced total of 20 exposures over period of 90 days under identical conditions. (B) Percentage of divers[a] who developed symptoms as a function of exposure to 100% oxygen under various pressures. Both plots were adapted from literature [5].

When these signs of toxicity occur, convulsions can be often [1] mitigated by lowering the $pO_2$ and/or the total pressure to a normal level, [1,4] but administering benzodiazepines [13] or anaesthetics [9] can help in some [13] cases as well. However, quickly lowering the total pressure increases the risk of barotrauma in patients (dysbarism) and gas embolism. [1] Because hyperbaric convulsions are very dramatic and potentially life-threatening, [1] current guidelines recommend avoiding pressures above 203 kPa during HBOT to mitigate the risk of hyperbaric convulsions as much as possible, [13,17] Nonetheless, limiting the total pressure decreases the $pO_2$ in tissues, [11] thereby lowering the efficacy of HBOT. [11] Thus, increasing the safety and efficacy of HBOT requires eliminating the risk of hyperbaric convulsions, [1] which has been difficult, because hyperbaric convulsions are still poorly understood. In this context, hyperbaric convulsions have been explained based on various hypotheses, most of which implicating ROS. [4,18,19]

*In vitro* [20–23] and *in vivo* [24–28] studies have shown that γ-aminobutyric acid receptors A (GABA$_A$) [20–23] and glycine [29,30] receptor function is altered under increased pressure. Moreover, the administration of anticonvulsants and sedatives can delay or even prevent the onset of hyperbaric

---

[a] Notoriously, divers are more susceptible to hyperbaric toxicity than patients in hyperbaric chambers.[5] The phenomenon is not well understood, but it may be partially attributed to increased oxygen consumption due to thermal losses caused by divers' contact with water.
[b] In extremely high $pO_2$ (600 to 1000 kPa), convulsions usually occur in mere minutes in mice[14,15] and humans[16]

convulsions, significantly prolonging the lifespan of laboratory animals in HBOT. [14,31] Based on mounting evidence, long overlooked, we postulate that hyperbaric convulsions may be also attributed to these pressure-induced reversible alterations in the function of neurotransmitter receptors.

In this evidence-based prospective review / hypothesis, we briefly examine previously proposed mechanisms of hyperbaric convulsions involving ROS and argue that other factors, primarily neurotransmitter receptor dysfunction, contribute to hyperbaric convulsions. We discuss the main findings that support our hypothesis. Lastly, we suggest administering anticonvulsants or sedatives before routine hyperbaric oxygen therapies and deep-sea diving to prevent hyperbaric convulsions.

## 1.1 ROS only partly account for hyperbaric convulsions

Hyperbaric convulsions are most often attributed to the toxic effects of ROS.[3,4,19,32] ROS encompass a group of highly reactive oxygen-containing compounds that can be formed *in vivo* and damage tissues. Indeed, ROS concentration in tissues increases with $pO_2$ during HBOT, [18] which in turn can damage organs such as lungs and retinas. [6,8] Thus, ROS is commonly accepted as a mechanism of peripheral oxygen toxicity. [6] In line with this hypothesis, antioxidants quench ROS and thus mitigate oxygen toxicity in peripheral organs. [6]

Moreover, ROS affect cellular receptors *via* covalent irreversible modifications, thereby altering neuronal function. [32–36] These irreversible modifications may be implicated in the CNS toxicity of HBOT, which would explain brain sensitisation to oxygen after prolonged or repeated HBOT. However, hyperbaric convulsions can be quickly mitigated by lowering the $pO_2$ to normal conditions, which shows that irreversible modifications cannot be the only causative factor of oxygen-induced convulsions. And while a few antioxidants (*e.g.* disulfiram, glutathione, and caffeine)[1,37] have been shown to prevent or mitigate hyperbaric convulsions,[c] many potent antioxidants (vitamin E,[39] *N*-acetylcystein, [39] allopurinol, [40] hypoxantine, [40] superoxide dismutase [1]) have no effect, suggesting that ROS may only play a minor role in the acute onset of hyperbaric convulsions. In other words, other mechanisms are needed to explain the acute CNS toxicity of HBOT.

## 1.2 Other theories suspect neurotransmitters may have a direct role

Some theories suspect that HBOT can cause imbalance of excitatory and inhibitory neurotransmitters, and thus lead to convulsions. In line with this theory, compounds that excite CNS (thyroid hormones, epinephrine, among others) increase the chances of convulsions. [1] Indeed, numerous studies confirm that HBOT decreases concentrations of γ-aminobutyric acid (GABA) in brain,[15,41–43] the main inhibitory neurotransmitter in the brain, while leaving concentrations of excitatory amino acids (glutamate and aspartate) unchanged. [41] This decrease of GABA is usually ascribed to HBOT-decreased brain activity of glutamate decarboxylase (GD; EC 4.1.1.15), [42–44] the main enzyme involved in synthesis of GABA, [43] Further confirming this hypothesis, when GD activity was blocked pharmacologically, laboratory animals were more prone to HBOT-induced convulsions.[45] This temporary decrease of GD activity during HBOT in brain is not fully understood, but is often attributed to ROS-induced deactivation of GD.[43,44] Noteworthily, administering pyridoxine (vitamin $B_6$; its active form pyridoxal phosphate is a co-factor of GD) accelerated normalization of GD activity, but it still takes *ca.* 1 to 2 hours in rats.[45] Therefore, HBOT-derived decrease of GD activity may offer a plausible explanation of hyperbaric convulsions and pyridoxine (or other forms of vitamin $B_6$) may help mitigate its consequences.

---

[c] Noteworthily, these drugs are likely to have their own pharmacologic effects, which counter the hyperbaric convulsions in with a ROS-independent mechanism, as has been speculated in previous studies.[38]

The return of GD activity to normalcy is relatively slow (> 1 hour in rats), [45][d] during which the animals may experience multiple seizures,[45] which is in contrast with the fact that hyperbaric convulsions usually cease within minutes after the decrease of pressure to normal levels. In other words, although HBOT-induced decrease of GD may contribute to the onset of hyperbaric oxygen convulsions, this mechanism does not explain why most hyperbaric oxygen convulsions are mitigated within minutes when $pO_2$ and/or pressure is returned to normalcy. Moreover, while some studies say that pyridoxin had protective role against HBOT-induced convulsions,[46] other studies do not support this claim.[47] Therefore, an additional mechanism is needed to explain this sudden stop of hyperbaric convulsions.

## 1.3 High pressure alters the function of receptors

Until now, theories explaining the mechanism of hyperbaric convulsions have overlooked that high pressure can alter the function of cell receptors. For example, increased ambient pressure can decrease sedative effect of many drugs, commonly known as "pressure reversal of anaesthesia". More specifically, increased pressure decreases the *in vivo* narcotic potency of many (but not all [48]) general anaesthetics (methoxyflurane,[49] ethanol,[27,49,50] chloroform,[49,51] diethyl ether,[49,51] hexobarbital,[48] and nitrous oxide,[49,51,52] among others [49,51,52]); tranquilizers (chlorpromazine [49] and droperidol [49]); local anaesthetics (lidocaine [49] and procaine [49]) and other drugs that depress CNS (diazepam [49] and morphine [49]). This effect mainly depends on the total pressure rather than $pO_2$ (but an effect of $pO_2$ was also described [27,48]) [50]. The pressure reversal of anaesthesia is likely caused by alterations in the function of glycine and $GABA_A$ receptors [29,30] (and likely other receptors too).

Both glycine and $GABA_A$ receptors are heavily regulated by both endogenous and exogenous allosteric modulators. [53–55] But increased pressure induces changes in $GABA_A$ and glycine receptors that decrease their sensitivity to many allosteric modulators (**Figure 2**). Although high pressure does not decrease GABA-mediated chloride currents (**Figure 2A**) [22], it does reverse the activity of both positive (ethanol,[22,56] pentobarbital,[20,56] diazepam,[21,22,24] flunitrazepam,[20,56] and zolpidem [21]), and negative (Ro15-4513,[21] and 6,7-dimethoxy-4-ethyl-β-carboline-3-carboxylate (DMCM) [21]) allosteric modulators, as shown in **Figure 2**. By contrast, some allosteric modulators are resistant to high-pressure effects on their activity. As a case in point, the activity of 3α-hydroxy-5β-pregnan-20-one is not neutralized under high pressure (**Figure 2D**). [56] As with $GABA_A$ modulators, the activity of positive allosteric modulators of glycine receptor (ethanol, [57,58] butanol, [58] and propofol [59]) can also be neutralized under high pressure. These pressure-induced changes in neurotransmitter receptor activity may trigger pressure reversal of anaesthesia.

---

[d] More than 1 hour even when pyridoxine is administered; under normal conditions, the recovery may be even longer, but the data is missing in literature

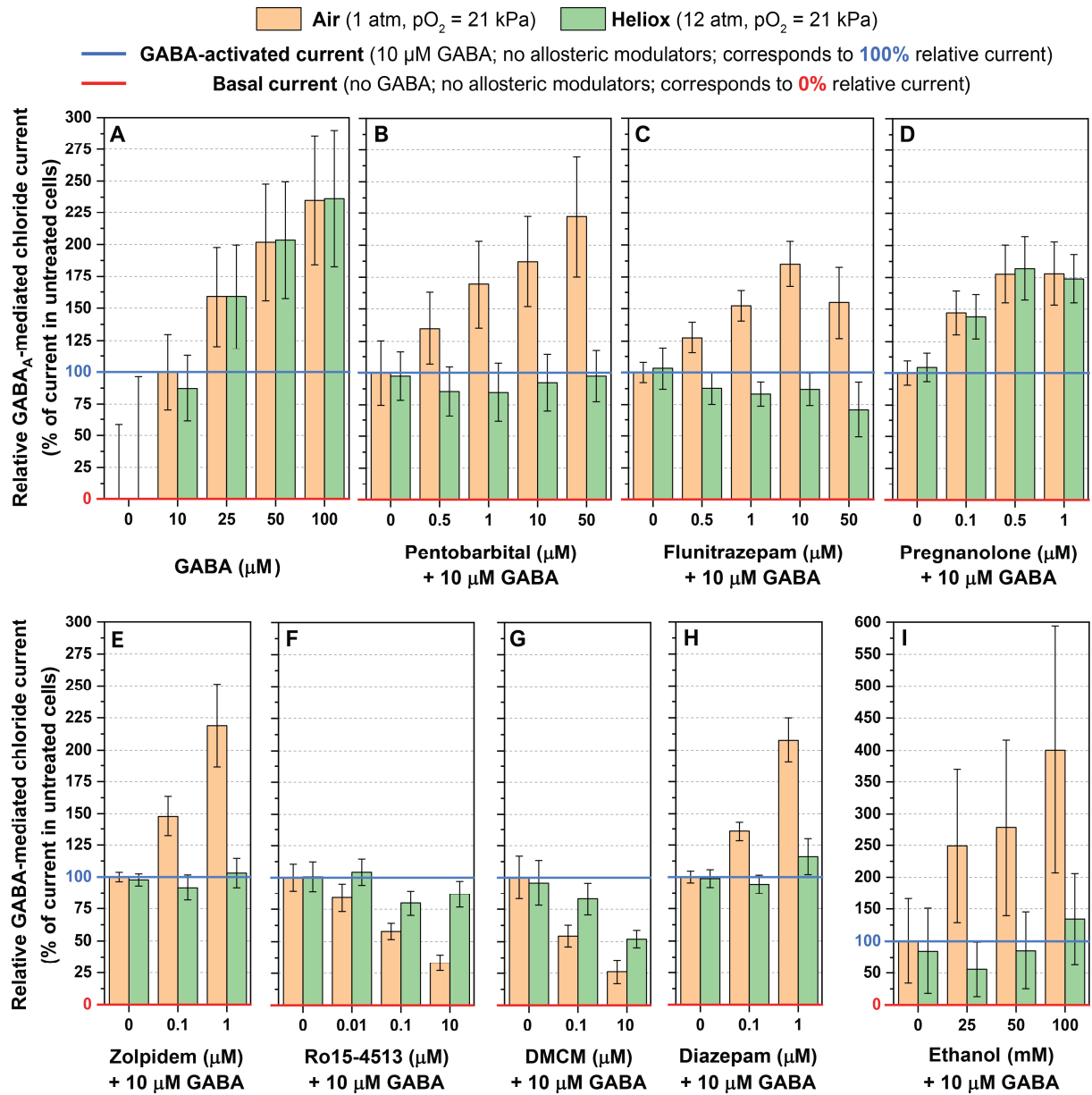

**Figure 2**. (**A**) Relative GABA-mediated chloride flow in brain tissues *in vitro* as a function of concentration of GABA. Effect of various positive (**B**, **C**, **D**, **E**, **H**, and **I**) and negative (**F** and **G**) allosteric modulators on GABA-mediated chloride currents in brain tissue *in vitro* under normal (air, 20.9% of $O_2$; pressure 1 atm.; $pO_2$ = 21 kPa) and high pressure (heliox mixture; 1.7% $O_2$, 98.3% He; pressure 12 atm.; $pO_2$ = 21 kPa). Data compiled from previous studies, [20–22] re-calculated, and expressed as mean ± standard error.

The mechanism how pressure affects GABA-mediated chloride flow remain unknown. [60] Original hypotheses assumed that pressure induces changes in receptors, which changes the receptors' affinity towards allosteric modulators, but these hypotheses have yet to be confirmed. [60] However, available data show that affinity of flunitrazepam towards GABA receptors is not affected by pressure. [20] Noteworthy, the effect of pressure on cell receptors does not have to affect receptors directly but may alter the surrounding structures of receptors and thereby affect the receptors. For example, caveolae rafts are pressure-sensitive [61,62] and ROS-sensitive [63] and are often associated with receptors (even G-protein coupled receptors),[64] whereby pressure can alter cell signalling and communication pathways [61,62]. Nevertheless, current data on effect of pressure on ionotropic receptors via caveolae rafts and other pressure-sensitive structures are very limited and should be investigated in future studies.

Regardless of the actual mechanism of pressure-derived decrease of GABA-mediated chloride flow, herein, we argue that pressure-reversal of anaesthesia and hyperbaric convulsions are two sides of one coin. In other words, these pressure-induced changes in the function of ionotropic receptors may explain not only pressure-reversal of anaesthetics (exogenous allosteric modulators) but also high-pressure-induced convulsions. Increased pressure induces changes in both $GABA_A$ and glycine receptors which, in turn, decrease the activity of endogenous (and exogenous) allosteric modulators. As a result, increased pressure weakens repolarization of neurons and ultimately increasing the risk of hyperbaric convulsions (**Figure 2**). In that sense, high pressure resembles both mechanism and clinical manifestations of $GABA_A$/ glycine receptor antagonists ("convulsants"), [65,66] and, therefore, may be treated similarly (*i.e.* administration of anaesthetics [67] or sedatives [68,69]). Considering this mechanism, administering exogenous positive allosteric modulators may prevent hyperbaric convulsions.

## 1.4 Convulsions result from multiple pathophysiological pathways

Convulsion that occurs during HBOT may be a consequence of multiple pathological pathways that have been affected by HBOT. Our proposed mechanism (1) of the direct effect of pressure on cell receptors does not exclude the remaining proposed mechanisms of hyperbaric convulsions, such as (2) ROS-induced changes in receptors or (3) decreased GABA synthesis but is just their extension. While the later two theories (2+3) can explain the long-term increasing chance of their occurrence with prolonged or frequent exposure to HBOT, they do not explain the sudden onset and termination of convulsions. In contrast, direct hyperbaric changes in the function of receptors (1) may explain the sudden onset and termination of convulsions with increasing and decreasing pressure. Nevertheless, mechanism (1) does not easily explain the increasing chances of hyperbaric convulsions with time of exposure, and frequency of HBOT, nor why increasing partial pressure of oxygen shortens the typical onset of convulsions. However, combining these three theories may explain most experimentally observed factors that affect the onset of hyperbaric convulsions (**Figure 3**). Moreover, this complexity of the mechanism may explain, why so many seemingly unrelated factors, and why many internal and external factors may increase or decrease the risk of convulsions, and why some studies come to conflicting conclusions about the efficacy of specific treatments (*e.g.*, [46] and [47]).

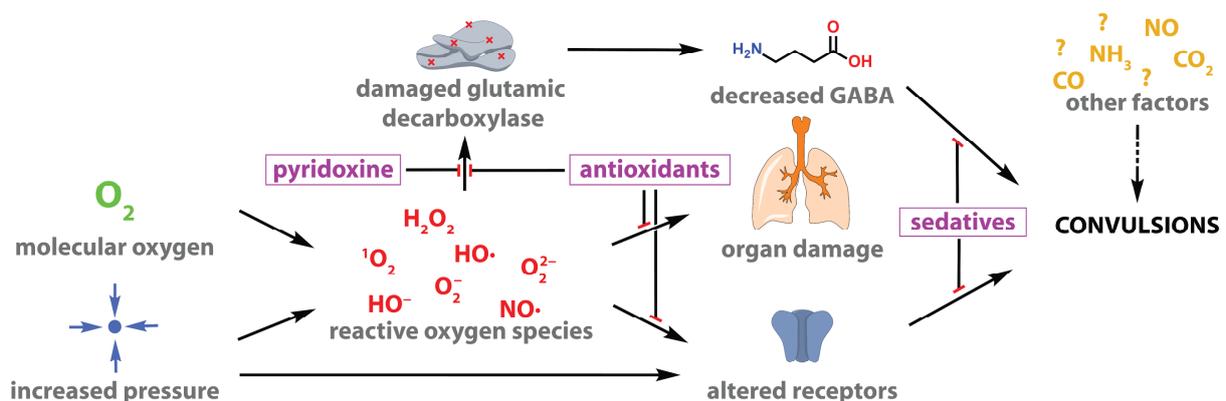

**Figure 3**. Mechanisms of hyperbaric toxicity. Increased pressure and oxygen concentrations enhance the production of ROS, which damage peripheral organs and alter cell receptors, irreversibly decreasing the threshold to convulsions. Moreover, ROS damages glutamic decarboxylase, thereby decreasing synthesis of GABA, further predisposing convulsions. Furthermore, pressure directly alters the functions of receptors, thereby reversibly facilitating convulsions. While the effect of ROS can be mitigated using antioxidants and pyridoxine, convulsions can be prevented with sedatives.

A few factors *e.g.*, carbon monoxide poisoning, exercise, hypercapnia, *etc.* increase the risk of convulsions), yet they are not explained by a combination of these three theories, suggesting that other mechanisms are yet to be described. Some studies show that ammonia [42] or nitric oxide [38,70,71] may be involved, which may alter the brain function in other ways (*e.g.*, NO-mediated transient increase of cerebral blood flow may predispose to convulsions [38,71]). In line with this hypothesis, compounds that affect brain blood flow (*e.g.*, acetazolamide [72]) have been reported to increase the risk of hyperbaric convulsions and shorten the animal's lifespan [72]. However, it is not clear whether these changes in concentrations of small molecules and overall changes are truly a cause, or just a consequence of underlying mechanism. Therefore, future research should enlighten the role of these endogenous molecules in hyperbaric convulsions towards finding the best preventive treatment.

## 2.1   Low doses of sedatives may prevent the onset of convulsions

In line with our hypothesis, phenobarbital and number of other barbiturates,[14,31] diazepam and other benzodiazepines,[14,31] chloroform,[9] baclofen,[31] carbamazepine,[73] chlorpromazine, [1] and antagonists of excitatory amino acids,[74] and vigabatrin [75] (anticonvulsive inhibitor of GABA degradation [76,77]) have been shown to delay or even prevent hyperbaric seizures and death in laboratory animals. Although these compounds strongly limit hyperbaric convulsions in animal models, they are not routinely administered before HBOT to prevent hyperbaric convulsions. In fact, this prevention strategy has been hinted as a possibility before,[7] but hardly ever used in clinical practice. Administering sedatives and anticonvulsants to test animals significantly prolongs their survival, [14,31] which indicates that these compounds truly prevent the convulsions rather than simply masking them. Accordingly, anticonvulsant medication may enable exposure to higher $pO_2$ during HBOT and/or decrease the necessary intervals between sessions, thus increasing HBOT efficacy. Moreover, low doses of anticonvulsants may be particularly beneficial to deep sea divers and patients with a high risk of convulsions (*e.g.*, carbon monoxide poisoning patients [12]). Therefore, administering these compounds may effectively mitigate hyperbaric-induced acute CNS toxicity.

## 2.2   Effective preventive therapies may encompass multiple compounds

A combination of sedatives, other drugs antioxidants, and pyridoxine may be the best preventive therapy to reduce the risk of hyperbaric convulsions. HBOT causes many alterations in the organism, which may ultimately lead to convulsions and peripheral toxicity (**Figure 3**). As discussed above, sedatives/anticonvulsants,[14,31] some antioxidants,[1,37] and pyridoxine [46] can prevent the onset of HBOT toxicity and/or lower its consequences in animals or humans, and therefore their preventive administration before HBOT should be considered. While these compounds may have a limited effect on their own, their concurrent administration may have a more profound protective effect. In addition to its toxic effects on CNS, HBOT may also damage peripheral tissues (*i.e.*, peripheral toxicity) as a result of prolonged exposure to ROS.[8,78] Nevertheless, peripheral oxygen toxicity can be supressed by administering various anti-oxidants.[8,18,78] Thus, sedatives or tranquilizers should be combined with strong antioxidants to prevent acute CNS toxicity and ROS-induced chronic peripheral damage, respectively.

## 3   Future perspectives

The lack of prevention strategies for hyperbaric convulsions hinders attempts to broaden the clinical applications of HBOT.[1,2] Therefore, future studies should determine whether the combination of antioxidants, pyridoxine, and sedatives may prevent the hyperbaric oxygen toxicity. Moreover, future studies should ascertain whether "pressure-sensitive" drugs that are antagonized under high pressures (*e.g.*, phenobarbital or diazepam[e]) are more effective than those that are not (*e.g.*, neurosteroids or gaboxadol [56]) and compare those with direct $GABA_A$ agonists (*e.g.*, progabide or γ-amino-β-hydroxybutyric acid) and indirect $GABA_A$ agonists (*e.g.*, vigabatrin). The efficacy of other inhibitory compounds that do not directly affect ionotropic $GABA_A$ and glycine receptors (*e.g.* baclofen, melatonin, ketamine, haloperidol, levomepromazine, fentanyl, or suvorexant), because the inhibitory effect of such compounds would be independent of the signalling pathways that are inhibited by high ambient pressure. Because even closely related compounds have shown vastly different properties in anaesthesia reversal and convulsion prevention (**Figure 2** and [20–22]), such studies must identify the best compounds for clinical applications. Moreover, future research should also aim at understanding how increased pressure affects receptors to highlight the underlying mechanism of HBOT toxicity and hence to avoid its adverse effects. Finally, future studies should elucidate why some factors (carbon monoxide, exercise, *etc*.) increase the chances of convulsions. Only by gaining deeper insights into the underlying causes of hyperbaric convulsions, we may be able to develop such prevention strategies.

## Conclusion

HBOT toxicity affects the entire body with several intertwined mechanisms. First, increased $pO_2$ increases the concentrations of ROS in tissues and thus may irreversibly alter the function of CNS, causing a long-term predisposition for convulsions. Second, increased ROS in brain reversibly decreases concentrations of GABA, further predisposing patients to convulsions. Third, increased ambient pressure alters the function of GABA and glycine receptors (phenomenon closely related to pressure-reversal of anaesthesia), which may cause acute and quickly reversible convulsions. Based on this complex mechanism, we suggest that concurrent administration of antioxidants (to neutralize ROS; discussed in *section 1.1*), pyridoxine (to boost synthesis of GABA; discussed in *section 1.2*), and GABA-active drugs (sedatives or anticonvulsants to reverse the pressure-inhibited $GABA_A$ and glycine receptors; discussed in *section 1.3*) may prevent the onset of hyperbaric oxygen toxicity. Such preventive strategies may be employed to improve the safety and efficacy of hyperbaric oxygen therapy and deep-sea diving.

## Acknowledgement

We would like to thank prof. Kai Simmons, Dr. Tomáš Slanina, MSc. David Dunlop, and MSc. Jan Kadlec for their critical comments and helpful suggestions. The authors also thank Dr. Carlos V. Melo for editing the manuscript.## Conflicts of interest

None to declare.

---

[e] The effect of phenobarbital and diazepam was reversed at 12 atm. of pressure, but lower pressures (closer to those used in HBOT, 2 to 3 atm.) may not be sufficient to fully reverse their effects.